\documentclass[aps,pra,twocolumn,groupedaddress]{revtex4-1}
\usepackage[latin1]{inputenc}
\usepackage{amsmath}
\usepackage{amsfonts}
\usepackage{amssymb}
\usepackage{graphicx}

\begin{document}
	
\title{Incoherent quantum feedback control of collective light scattering by Bose-Einstein condensates}
\author{Denis A. Ivanov}
\address{St.~Petersburg State University,\\
Ulianovskaya 3, Petrodvorets, St. Petersburg, 198504, Russia}
\author{Tatiana Yu. Ivanova}
\address{St.~Petersburg State University,\\
Ulianovskaya 3, Petrodvorets, St. Petersburg, 198504, Russia}
\author{Igor B. Mekhov}
\address{University of Oxford, Department of Physics,\\
Clarendon Laboratory, Parks Road, Oxford OX1 3PU, UK}

\begin{abstract}
It is well known that in the presence of a ring cavity the light scattering from a uniform atomic ensemble can become unstable resulting in the collective atomic recoil lasing. This is the result of a positive feedback due to the cavity. We propose to add an additional electronic feedback loop based on the photodetection of the scattered light. The advantage is a great flexibility in choosing the feedback algorithm, since manipulations with electric signals are very well developed. In this paper we address the application of such a feedback to atoms in the Bose-Einstein condensed state and explore the quantum noise due to the incoherent feedback action. We show that although the feedback based on the photodetection does not change the local stability of the initial uniform distribution with respect to small disturbances, it reduces the region of attraction of the uniform equilibrium. The feedback-induced nonlinearity enables quantum fluctuations to bring the system out of the stability region and cause an exponential growth even if the uniform state is globally stable without the feedback. Using numerical solution of the feedback master equation we show that there is no feedback-induced noise in the quadratures of the excited atomic and light modes. The feedback loop, however, introduces additional noise into the number of quanta of these modes. Importantly, the feedback opens an opportunity to position the modulated BEC inside a cavity as well as tune the phase of scattered light. This can find applications in precision measurements and quantum simulations.
\end{abstract}

% Uncomment for keywords
%\vspace{2pc}
%\noindent{\it Keywords}: XXXXXX, YYYYYYYY, ZZZZZZZZZ
%\submitto{\LP}
% Uncomment if a separate title page is required
%\maketitle
% 
% For two-column output uncomment the next line and choose [10pt] rather than [12pt] in the \documentclass declaration
%\ioptwocol
%

\maketitle

\section{Introduction}
Feedback control is known to be a useful tool to modify dynamics of classical systems~\cite{fb-class}. In resent years the application of feedback to quantum systems attracts considerable attention. Since the appearance of the elaborated theory due to Wiseman and Milburn~\cite{wiseman-milburn,wiseman} there were many theoretical proposals~\cite{tombesi,jacobs} and experimental realizations~\cite{rempe,morrow} of feedback control in optics and atomic physics. Among them many feedback-based cooling methods for atoms, molecules and opto-mechanical systems have been discussed~\cite{raizen,wallentowitz,averbukh}. Most of these discussions involve a negative feedback loop as a mechanism to compensate for the thermal noise. However there are circumstances where the positive feedback can improve the performance of an atomic or molecular system. One possibility is to enhance the cooling and the self-organization in ensembles of particles coupled to a cavity field. For review of these systems without feedback see~\cite{cav-cool-rep,rev-mekhov-ritsch}. A semi-classical considerations of the application of feedback are given in Ref.~\cite{jphysb} for cooling and in Refs~\cite{ivanov-jetp-lett,ivanov-jetp} for the self-organization~\cite{so-bec1,so-bec2,so-bec3}. 

In this paper we consider the application of feedback to a Bose-Einstein condensate (BEC) in a configuration similar to a collective atomic recoil laser (CARL)~\cite{carl-therm,carl-bec1,carl-bec2}. A CARL setup consists of atoms inside a ring cavity. The cavity is pumped so that only a single running mode is excited. The uniform distribution of atoms for certain pump becomes unstable. The fluctuation of atomic densities result in scattering of light into the mode running in the opposite direction with respect to the pumped mode. As a result of the interference between the pumped and the scattered modes a standing wave optical potential is generated that makes further redistribution of the atoms favorable. Finally the atoms are redistributed in a Bragg grating that effectively scatter the pumped light and amplify the counterpropagating mode. This is similar to a laser. 

The CARL effect appears as a result of positive feedback due to the ring cavity. The cavity performs the feedback that is proportional to the integrated signal over some time interval defined by the cavity photon lifetime. One, however, can imagine a more elaborated feedback algorithm such as  proportional - integral - derivative control (PID-control)~\cite{fb-class}, that can lead to a superior performance or even introduce some new features into the system dynamics. The simplest possibility is to use an electronic circuit that processes the photocurrent produced by the scattered light. In the limit of cold particles the bandwidth of typical electronic components should be wide enough to correctly process the photodetected signal. The feedback mechanism is based on the extraction and procession of a classical signal, which introduces the measurement backaction noise into the dynamics. The effect of this noise is one of the main subjects of this paper.

Also in the absence of the feedback the phase of the scattered CARL light and the position of the atomic Brag-grating are not defined. In the feedback scheme, based on the photocurrent measurement of the scattered light, the position of the generated periodic pattern of the atoms as well as the phase of the scattered light will be locked to the potential that serves as the feedback actuator. 

Such positioning of atoms and control of quantum light can assist in quantum simulations, based on ultracold gases inside cavities \cite{mekhov1,mekhov2,mekhov3,mekhov4}, where even small shift in the position leads to very different quantum phases. This is in line with recent breakthrough experiments, where an external optical lattice trapping quantum gases was implemented inside a cavity for the first time~\cite{hemmerich,esslinger}. These experiments open the field of cavity QED for truly many-body strongly correlated states of quantum matter.

We will explore the ultimate quantum regime where the number of excitations in BEC and the number of photons are small, so the quantum theory of operation of the feedback loop~\cite{wiseman} will be eventually applied. Using quantum trajectories~\cite{quant-traj-1,quant-traj-2} as well as semi-classical equations we solve for the dynamics of the system. It appears that the feedback loop makes the stability of the system dependent on the value of the fluctuations. The quantum solution demonstrates that the fluctuations in the number of BEC excitations grow with the value of the feedback strength. This is the result of the amplification of the photon number noise by the feedback loop.     

The paper is organized as follows. The description of the physical model and the mathematics used for it are given in Sec.~\ref{sec:model}. The results of semi-classical and fully quantum treatment are presented and discussed in Sec.~\ref{sec:cq-dynamics}. The main conclusions of the paper are given in Sec.~\ref{sec:conclusion}. 

\section{Model}
\label{sec:model}
\begin{figure}
\includegraphics[width=1.0\linewidth]{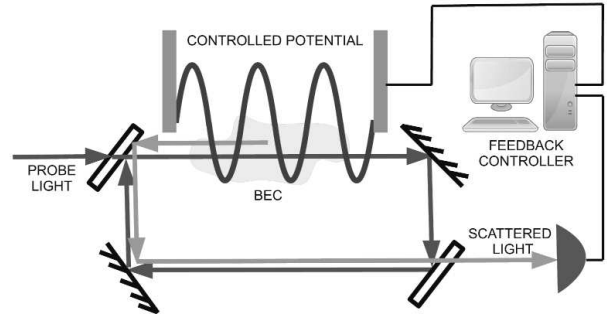}
\caption{Principal scheme of the setup. Atomic BEC is inside the ring cavity. The light scattered by BEC form the probe beam is detected and used to control additional potential via an electronic circuit.}
\label{fig:fig1}
\end{figure}

Although the method is potentially applicable to all polarizable particles including molecules~\cite{mekhov1-LP13}, we assume that the particle ensemble is represented by trapped atoms cooled below the Bose-Einstein condensation temperature. The atom-atom interactions will be neglected since quite dilute atomic ensembles, $N \!\sim\! 10^4$ atoms, will be considered. The atoms are placed inside a ring optical cavity, see Fig.~\ref{fig:fig1}, and interact with it scattering photons from the externally pumped running mode into the counter-propagating one.
We restrict the consideration to only two counter-propagating modes of the same frequency, which is assumed to be well red-detuned from the atomic transition. Moreover, as we are interested in the evolution of the atomic sub-system and its ability to form periodic density patterns, we consider regimes where the field of the coherently pumped mode $\eta$ can be considered constant. Thus only one operator cavity field $a$ remains for the quantum dynamical problem. Adiabatic elimination of the excited atomic state gives the following result for the atom-field Hamiltonian
\begin{eqnarray}
\label{eq:H0}
&&H_\mathrm{0} = \hbar \Delta a^\dagger a + \int dx \psi^\dagger(x)\Big[-\frac{\hbar^2}{2 m} \partial_x^2 
\\
&+& \hbar U_0 \left(a^\dagger a + |\eta|^2 + a^\dagger \eta e^{2ikx} + \eta^* a e^{-2ikx}\right)\Big]\psi(x), \nonumber
\end{eqnarray}
where $\psi(x)$ is the bosonic field operator obeying $\left[\psi(x),\psi^\dagger(x')\right] \!=\! \delta(x-x')$, $\Delta$ is the atom-field detuning, $U_\mathrm{0}$ is the atom-field interaction constant, and $m$ is the mass of the atom. Apart from the mentioned above simplified consideration of the cavity, this is the well-known result that can be found for example in Ref~\cite{meystre}. 

Together with the cavity field the controlled potential is applied to the atoms. The control signal determines the depth of this potential. The signal is derived from the result of the measurement of the number of photons leaving the scattered cavity mode. In principle, one can design a control scheme where the potential depth would be any function of the detected signal. We assume the linear feedback control as it will be shown to be enough for our purpose. This is described by the following additional term to the system Hamiltonian~(\ref{eq:H0})
\begin{equation}
\label{eq:Hfb}
H_\mathrm{fb} = i(t- \tau) Z.
\end{equation}
Here $i(t-\tau)$ is the photocurrent detected outside the cavity. The description of feedback can take into account possible delay of the electronic feedback loop, which is represented by the time interval $\tau$ in Eq.~(\ref{eq:Hfb}). Here we assume that the photocurrent $i(t)$ is a classical quantity that represents the output of the classical apparatus. The feedback operator $Z$ in our case acts only on the atomic subspace of the total Hilbert space, it represents the additional controlled potential for the atoms. It is assumed to be the standing optical wave with the well defined phase
\begin{equation}
\label{eq:Z}
Z = U_0 K \int dx \psi^\dagger(x) \sin^2(kx) \psi(x).
\end{equation}
The parameter $K$ is the feedback gain, it is defined in such a way that the intensity of the lasers forming the feedback potential is given by $I_\mathrm{fb}(t) \!=\! K i(t - \tau)$. The atom-field coupling parameter $U_0$ here is assumed to be the same as in Eq.~(\ref{eq:H0}) since the similar frequencies should be used for the probe and feedback light fields. 

Note that an interesting generalization of the described feedback setup exists. For the feedback loop one can use light scattered at any angle with respect to the BEC axis and the pump beam direction. This light contains different information about the distribution of the atoms~\cite{atom-qnd1,atom-qnd2,atom-qnd3} than the back scattered light. Thus generation of more exotic atomic density distributions and BEC states may be possible.  

To clarify the sort of feedback we are dealing with, we distinguish between two types of photo-detectors. The first type is a very fast device that can reproduce all features of the system evolution and response to a single photon. In this case the photo-current is given as, $i(t) = dN(t)/dt$, the derivative of the total number $N(t)$ of photo-counts, which is a Poisson stochastic process, having the ensemble average $E(dN(t)) \!=\! 0$ and the moment $E(dN(t) dN(t')) \!=\! \langle a^\dagger a \rangle \delta(t-t')$. It has been shown by Wiseman~\cite{wiseman} that in order to correctly handle this type of photo-detection one can use the master equation that in our notation can be written as
\begin{equation}
\label{eq:ME}
\frac{d \varrho}{dt} = -\frac{i}{\hbar}\left[H_\mathrm{0}, \varrho \right] + \kappa\mathcal{D}\left[ae^{-i Z}\right] \varrho,
\end{equation}    
where the decay superoperator reads as $\mathcal{D}[c]\varrho \!=\! c\varrho c^\dagger -(c^\dagger c \varrho + \varrho c^\dagger c)/2$. The parameter $\kappa$ is the cavity photon decay rate.
This equation governs the evolution of the unconditioned density matrix of the whole system, BEC and the cavity field. The equation~(\ref{eq:ME}) is valid in the Markovian limit, that is neglecting the time delay $\tau$. We assume that in the considered system the Markovian limit can be at least potentially satisfied since the feedback circuit operates with electro-magnetic signals which are much faster than the motion of atoms. 

In contrast to the described above, the other type of feedback uses an integrating detector that averages the measurements over some time interval $T$. If this interval is small compared with the atomic evolution time scale and simultaneously large compared with the photon cavity decay rate then it is natural to approximately consider the time average over $T$ as an ensemble average. The photo current is then given by $i(t) = \kappa\mathrm{Tr}\left\{\varrho(t) a^\dagger a \right\}$. Inserting this into Eq.~(\ref{eq:Z}) one ends with a non-linear master equation. It is rather straightforward to check that the corresponding equations for average values will contain the feedback contribution as the potential for atoms proportional to the scattered photon number. No feedback induced higher-order correlations appear. Such a feedback effect agrees with the classical intuition and represents the physical model of the semi-classical approximation that we will also address below.

We start with the first type of feedback that preserves quantum correlations. Following the convenient approach developed in Ref.~\cite{meystre} we decompose the atomic field into a cosine and sine excitation modes as
\begin{equation}
\label{eq:3-mode-def}
\psi(x) = \frac{1}{\sqrt{L}} \psi_\mathrm{0} + \sqrt{\frac{2}{L}} \psi_\mathrm{c} \cos(2kx) + \sqrt{\frac{2}{L}} \psi_\mathrm{s} \sin(2kx).
\end{equation}
Here $L$ is the length of the BEC sample, the bosonic annihilation operators $\psi_\mathrm{0}$, $\psi_\mathrm{c}$, and $\psi_\mathrm{s}$ describe uniform distribution of atoms, their cosine and sine modes, respectively. We assume that the number of atoms in the uniform BEC $N$ is large enough so that it only negligibly changes due to the interaction with the cavity field. The large $N$ also allows to neglect the operator character of $\psi_\mathrm{0}$ and consider $\psi_\mathrm{0} \!=\! \sqrt{N}$.  

Using the 3-mode representation of the atomic field, Eq.(\ref{eq:3-mode-def}), one can rewrite the Hamiltonian~(\ref{eq:H0}) and the master equation in a simpler form. This form of the master equation will be used for the direct numerical simulations. Having this approximation one obtains the following Hamiltonian 
\begin{eqnarray}
\label{eq:3-mode-H}
H_\mathrm{0} &=& 4\hbar\omega_\mathrm{R} \left(\psi_\mathrm{c}^\dagger \psi_\mathrm{c} + \psi_\mathrm{s}^\dagger \psi_\mathrm{s}\right) + \hbar U_\mathrm{0}\sqrt{N}\eta\left(a+a^\dagger\right)\left(\psi_\mathrm{c}^\dagger + \psi_\mathrm{c}\right) 
\nonumber \\
&+& i\hbar U_\mathrm{0}\sqrt{N} \eta \left(a -a^\dagger\right)\left(\psi_\mathrm{s}^\dagger + \psi_\mathrm{s}\right) .
\end{eqnarray}
Here we also define the recoil frequency $\omega_\mathrm{R} \!=\! \hbar k^2/2m$. To eliminate additional phase shift of the optical fields we assume the atom-field detuning to be given by $\Delta \!=\! U_\mathrm{0} N$. The action of the feedback potential $Z$ can also be rewritten using Eq.~(\ref{eq:3-mode-def}) to finally give 
\begin{equation}
\label{eq:feedback-shift}
e^{i\sqrt{\frac{N}{2}} K U_\mathrm{0}(\psi_c^\dagger + \psi_c)} \psi_c e^{- i\sqrt{\frac{N}{2}} K U_\mathrm{0}(\psi_c^\dagger + \psi_c )} = \psi_c - i\sqrt{\frac{N}{2}} K U_\mathrm{0} .
\end{equation}
The expression~(\ref{eq:feedback-shift}) is valid for $\omega_\mathrm{R}\gg U_\mathrm{0}$, otherwise it will contain addition phase multipliers. This condition can easily be fulfilled as we assume quite weak coupling of individual atoms to the cavity field. At the same time this condition does not mean that the evolution of the atoms is solely determined by $\omega_\mathrm{R}$ since the collective coupling $\sim U_\mathrm{0} \sqrt{N}$ still can be comparable or larger than the recoil frequency. To take into account the losses of atoms from the excited cosine/sine modes we add to the final version of the master equation~(\ref{eq:ME}) the dissipation terms $\gamma\mathcal{D}\left[\psi_\mathrm{c}\right] \!+\! \gamma\mathcal{D}\left[\psi_\mathrm{s}\right]$. The losses are assumed to have equal rate $\gamma$. Here we introduce the atomic losses phenomenologically and avoid discussing their physical mechanisms. For the detailed analysis of this aspect refer~\cite{domokos-losses}.  

\section{Classical and quantum dynamics}
\label{sec:cq-dynamics}

Before doing quantum trajectory simulations which are limited in the number of the excitation quanta, it is worth trying a semi-classical analysis. It should allow to test a broader range of parameters and approximately determine different regimes of the feedback operation. To do so we calculate the equations of motion for the average atomic and light fields using the master equation~(\ref{eq:ME}). It is convenient to write down these equations in terms of quadratures, defined as $x \!=\! (a + a^\dagger)/2$, $y \!=\! (a - a^\dagger)/(2i)$ for the cavity field and analogously for $X_\mathrm{c}$, $Y_\mathrm{c}$ and  $X_\mathrm{s}$, $Y_\mathrm{s}$ being the quadratures of the cosine and sine atomic modes, respectively. The resulting evolution equations read
\begin{eqnarray}
\label{eq:aver-evol}
\dot{x} &=& -2U_\mathrm{0}\sqrt{N}\eta X_\mathrm{s} - \frac{\kappa}{2}x, \nonumber\\
\dot{y} &=& -2U_\mathrm{0}\sqrt{N}\eta X_\mathrm{c} - \frac{\kappa}{2}y, \nonumber\\
\dot{X}_\mathrm{c} &=& 4\omega_\mathrm{R} Y_\mathrm{c} - \frac{\gamma}{2} X_\mathrm{c}, \nonumber\\
\dot{Y}_\mathrm{c} &=& -4\omega_\mathrm{R} X_\mathrm{c} - \frac{\gamma}{2} Y_\mathrm{c} - 2U_\mathrm{0}\sqrt{N}\eta x + \kappa K U_\mathrm{0}\sqrt{\frac{N}{2}}(x^2 + y^2) , \nonumber\\
\dot{X}_\mathrm{s} &=& 4\omega_\mathrm{R} Y_\mathrm{s} - \frac{\gamma}{2} X_\mathrm{s}, \nonumber\\
\dot{Y}_\mathrm{s} &=& -4\omega_\mathrm{R} X_\mathrm{s} - \frac{\gamma}{2} Y_\mathrm{s} - 2U_\mathrm{0}\sqrt{N}\eta y .
\end{eqnarray} 
Here the feedback enters via the last term in the equation for $Y_\mathrm{c}$ which is proportional to the feedback constant $K$. To obtain this last term it has been assumed that the average of the intracavity photon number can be approximated as the product of the average fields, $\langle a^\dagger a\rangle \!=\! \langle a^\dagger\rangle\langle a \rangle$. This is true for coherent cavity field states, but wrong for a general quantum state.

The easily seen steady state solution of Eqs~(\ref{eq:aver-evol}) is zero field in all of the involved modes. This solution is stable for small value of the pumping $\eta$, but at some threshold value of $\eta$ it loses stability and fluctuations grow to infinity. The unphysical infinity is the drawback of the model and is connected to the fact that there is no depletion for BEC and the cavity pump. Thus the considered model correctly describes the growth of the excitations only in the beginning of the evolution.  

\begin{figure}
\includegraphics[width=0.9\linewidth]{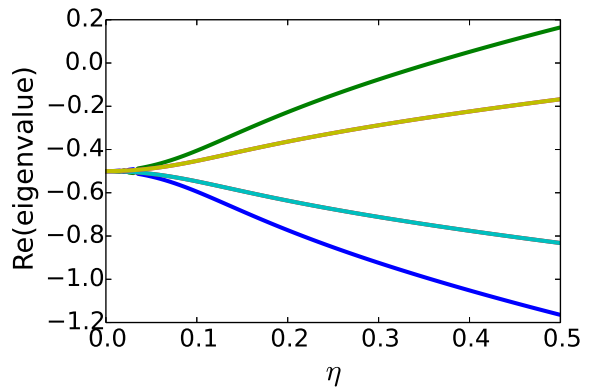}
\caption{Real parts of the eigenvalues of the semi-classical evolution, given by Eqs~(\ref{eq:aver-evol}), as functions of the pumping $\eta$. The parameters of the system are $U_\mathrm{0} \!=\! 0.1\omega_\mathrm{R}$, $N \!=\! 10^4$, $\kappa \!=\! 10\omega_\mathrm{R}$, $\gamma \!=\! 10\omega_\mathrm{R}$.}
\label{fig:fig2}
\end{figure}

The stability conditions for the zero steady state of Eqs~(\ref{eq:aver-evol}) can be analytically calculated but appear to be quite lengthy. It is much easier to check the stability for some numerical values. In all the simulations below we will use the following values $U_\mathrm{0} \!=\! 0.1 \omega_\mathrm{R}$, $N \!=\! 10^4$, $\kappa \!=\! 10\omega_\mathrm{R}$, $\gamma \!=\! 10\omega_\mathrm{R}$. These values correspond to quite weak coupling and moderate cavity finesse with respect to realistic cavity QED experiment with cold Rubidium~\cite{rb-cqed-exp}. The real parts of the eigenvalues of the evolution matrix corresponding to Eqs~(\ref{eq:aver-evol}) are shown in Fig.~\ref{fig:fig2}. 
It is seen that for some value of the pump field $\eta\! \approx\! 0.3$ the real part of one of the eigenvalues becomes positive and the zero steady state becomes unstable. Note that the presence of feedback does not affect the linear stability condition of the uniform equilibrium state as it is represented via a quadratic term and therefore vanishes in the linear approximation around the initial equilibrium with zero reflected field and atomic excitations.

\begin{figure}
\includegraphics[width=1.0\linewidth]{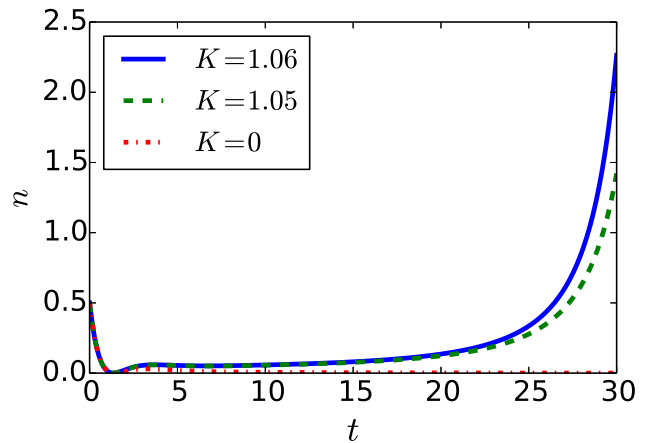}
\caption{The number of photons scattered from the atoms as a function of time for different values of the feedback constant $K$. The solid line corresponds to $K\!=\!1.06$, dashed line corresponds to $K\!=\!1.05$, dashed-dotted line corresponds to $K\!=\!0$. The pump is set below the threshold, $\eta \!=\!0.3$. The time scale is $\kappa^{-1}$.}
\label{fig:fig3}
\end{figure}

The presence of feedback however changes the range of attraction of the zero-reflection uniform equilibrium state, so that for relatively large values of the feedback constant $K$ even small fluctuations can force the system to loose its stability. This effect is demonstrated in Fig.~\ref{fig:fig3} where the number of photons $n$ in the scattered mode as a function of time is shown for several values of the feedback constant $K$. The cavity pump $\eta$ is assumed to be below the threshold. The dash-dotted line, that corresponds to $K\!=\!0$ thus approaches zero demonstrating the decay of initial fluctuations. The values of the initial fluctuations are set to quadrature variances of the vacuum state: $x(0) \!=\!y(0) \!=\!X_\mathrm{c}(0) \!=\!Y_\mathrm{c} \!=\!X_\mathrm{s} \!=\! Y_\mathrm{s} \!=\! 0.5$. For the value of $K \!=\! 1.05$ and the used initial fluctuations the system is unstable and the rapid growth of the scattered photons $n$ appears, see the dashed line in Fig.~\ref{fig:fig3}. The rate of this growth is very sensitive to the value of $K$, as it is seen from the comparison of the dash line ($K\!=\! 1.05$) and the solid line ($K\!=\!1.06$). Thus, though the feedback based on the photodetection does not change the linear stability of the zero steady state, in the presence of (quantum) fluctuations the feedback-induced nonlinearity can render the system unstable.

\begin{figure}
\includegraphics[width=1.0\linewidth]{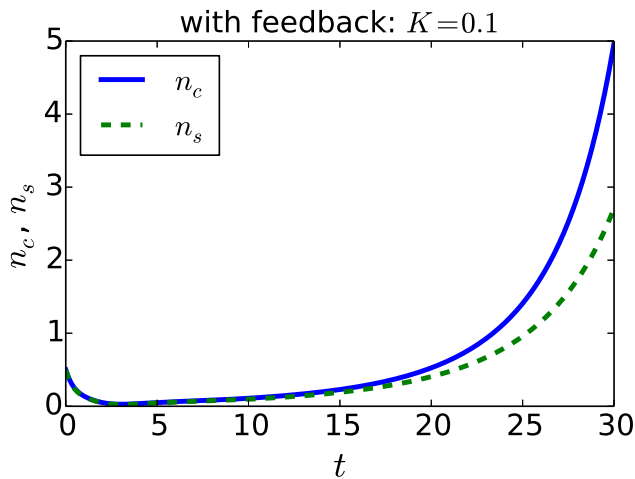}
\caption{The number of atoms in the "cosine" $n_\mathrm{c}$ and "sine" $n_\mathrm{s}$ modes of BEC. The pump is above the threshold, $\eta \!=\! 0.4$, the feedback constant is small, $K\!=\!0.1$. The time scale is $\kappa^{-1}$.}
\label{fig:fig4}
\end{figure}

The key (and the most promising for applications) difference between the system with and without the feedback is the possibility to control the position of the atomic Bragg grating and the phase of the scattered light. The solutions of the classical equations~(\ref{eq:aver-evol}) confirm the ability of the feedback loop to fix the position of the atomic density grating. The number of atoms in the "cosine" and "sine" modes of BEC with small feedback, $K\!=\!0.1$, are shown in Fig.~\ref{fig:fig4}. It is seen that the feedback discriminates the "sine" mode with respect to the "cosine" one, since the number of atoms in the "cosine" mode grows much faster which eventually results in the atomic distribution compatible with the feedback potential $Z$. As a consequence the phase of the light scattered from such an atomic Bragg-grating will be locked to some certain value. In the absence of the feedback there will be no predefined position of the atomic density grating, its position will be random according to initial density fluctuations. 

Note that this locking effect cannot be achieved by simply imposing some fixed external potential. This potential will lead to the oscillations in the number of atoms in the excited BEC modes without clear discrimination of one of them. Similar effect can be reached if one could control the losses of the excited atomic modes $\gamma$ and make them different for the "sine" and "cosine" modes. This could be possible, but the physical implementation of such an approach is not obvious.

Let us now address the quantum behavior of the system with and without the feedback. We solve the master equation~(\ref{eq:ME}) numerically using the method of quantum trajectories~\cite{quant-traj-1,quant-traj-2}. The simulation has been done with the Python package QuTiP~\cite{qutip}.

\begin{figure}
\includegraphics[width=1.0\linewidth]{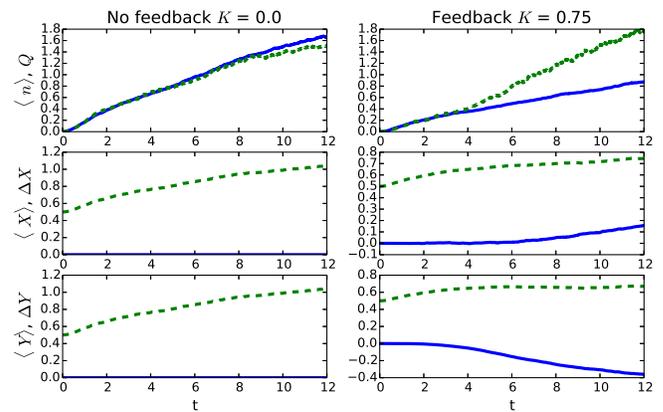}
\caption{The average values (solid lines) and variances (dashed lines) of the number of quanta and the quadratures of the scattered light. The left and right columns correspond to the case without and with the feedback, respectively. The time scale is $\kappa^{-1}$.}
\label{fig:fig5}
\end{figure}

In Fig.~\ref{fig:fig5} the result of the quantum simulation for the light mode is shown. The left column corresponds to the case above the threshold $\eta \!=\! 0.4$, but without the feedback, whereas the right column corresponds to some feedback $K \!=\! 0.75$ acting on the system below the threshold $\eta \!=\! 0.3$. The upper row shows the average number of the scattered photons $\langle n \rangle$ (solid lines) and the Mandel parameter $Q$ (dashed line). The second row shows the average $\langle X \rangle$ (solid line) and the variance $\Delta X$ (dashed line) of the amplitude quadrature. Finally, the third row corresponds to the phase quadrature $\langle Y \rangle$ and its variance $\Delta Y$.  The case without the feedback demonstrates that the light scattered from the BEC is in the thermal state as it has no well defined phase and the Mandel parameter equals the average number of photons.

The presence of feedback (left column in Fig.~\ref{fig:fig5}) results in the appearance of some additional noise in the number of photons, so that the photon number fluctuations exceed that of a thermal light. On the contrary, the fluctuations of the quadratures are of the same order as in the case without the feedback. Moreover, the feedback results in the appearance of nonzero values of the quadrature averages, which shows the existence of a well-defined phase in the scattered light.

\begin{figure}
\includegraphics[width=1.0\linewidth]{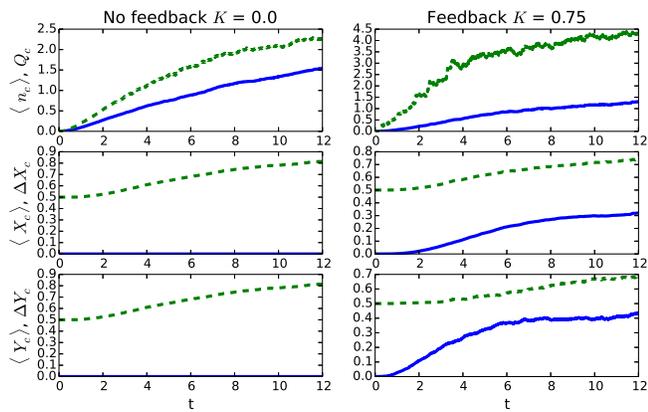}
\caption{The average values (solid lines) and variances (dashed lines) of the number of quanta and the quadratures of the excited atomic "cosine" mode. The left and right columns correspond to the case without and with the feedback, respectively. The time scale is $\kappa^{-1}$.}
\label{fig:fig6}
\end{figure}

What is more interesting is the quantum fluctuations of the excited atomic modes. The comparison between the averages and fluctuations of atom number and the quadratures for the "cosine" mode with and without the feedback is shown in the Fig.~\ref{fig:fig6}. It is worth noting that even without the feedback the fluctuations of the atom number are larger than in the thermal state, which is justified by the fact that the Mandel parameter $Q_c$ is larger than the average number of atoms $\langle n_c \rangle$. The "sine" and "cosine" modes interact with the light independently and do not directly interact with each other as we neglect the atom-atom scattering. Thus there is an uncertainty connected with the fact that it is not known which, "sine" or "cosine", was involved in a particular act of the exchange of quanta with the light field. This uncertainty explains the additional noise in the number of atoms of the atomic modes.

The feedback, in addition, induces the phase locking of the "cosine" mode, as nonzero average values of the $X_c$ and $Y_c$ quadratures appear. Similar to the photon number in Fig~\ref{fig:fig5} the fluctuations of the quadratures are not affected by the feedback loop and have approximately the same level as in the absence of the feedback. The atom number fluctuations of the "cosine" mode are increased by the feedback, which is similar to the mode representing the scattered light. 

\begin{figure}
\includegraphics[width=1.0\linewidth]{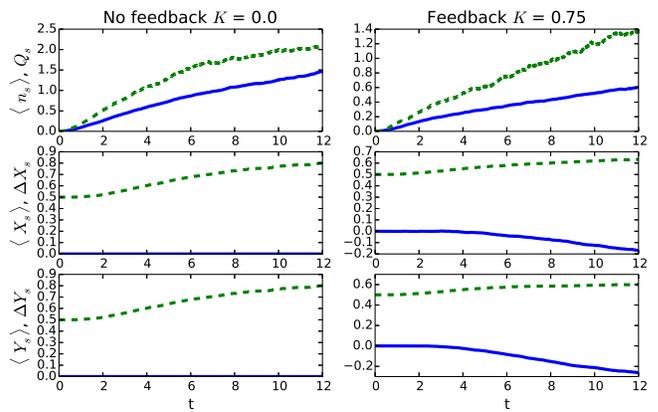}
\caption{The average values (solid lines) and variances (dashed lines) of the number of quanta and the quadratures of the excited atomic "sine" mode. The left and right columns correspond to the case without and with the feedback, respectively. The time scale is $\kappa^{-1}$.}
\label{fig:fig7}
\end{figure}

It is left to consider the behavior of the atomic "sine" mode. The corresponding averages and the fluctuations are presented in Fig.~\ref{fig:fig7}. The results corresponding to the case without the feedback coincide with the results for the "cosine" mode within the accuracy of the simulations, left column in Fig.~\ref{fig:fig6}. However, the results corresponding to the system with the feedback are quite different from that for the "cosine" mode. The fact that the feedback action is performed by the optical potential that is favorable for the "cosine" mode results in smaller population of the "sine" mode as it is seen from the comparison of the Figs~\ref{fig:fig6} and \ref{fig:fig7}. Thus the resulting density distribution of the atoms in BEC will eventually comply with the feedback potential which is predefined.

It is important to note that the feedback action almost does not affect the quadrature fluctuations of all the modes, even when the nonzero average values of the quadratures appear. The quantity that suffers from the additional noise due to the feedback is the number of quanta. This can be expected since the number of photons is detected and its noise is amplified in the positive feedback loop. Additional increase in the variances of atom numbers in the BEC modes is due to the fact that the light mode is separately coupled to the "sine" and "cosine" atomic modes. Thus if a photon appears in the light mode it remains uncertain which atomic mode is responsible for that.   

\section{Conclusion}
\label{sec:conclusion}
The application of the positive feedback loop to BEC with the assistance of a ring cavity has been considered. The probe light scattered from BEC into a cavity mode is detected and the signal is used to control the potential. The system is treated both semi-classically and fully quantum mechanically. The linear stability analysis of a semi-classical model shows that the feedback based on the photodetection does not change the stability of the zero-field equilibrium. However, numerical simulations show that the phase-space region of attraction of the zero-field equilibrium shrinks due to the feedback-induced nonlinearity. Thus the system may exhibit exponential growth if due to initial fluctuations it starts from some point in the phase space that is sufficiently remote from the equilibrium. In addition the feedback induces a predefined atomic density distribution and as a consequence a fixed phase of the scattered light.

The quantum treatment of the feedback was based on the Markovian approximation expressed by the master equation of Ref.~\cite{wiseman}. The numerical solution of this equation has shown that in the presence of feedback the quantum fluctuations can lead to the exponential growth even below the classical instability threshold. The fluctuations of the quadratures of all the modes are not increased by the presence of the feedback, which is not the case for the fluctuations of the number of quanta. The photon number fluctuations in the scattered light become stronger than in a thermal state due the noise amplification by the feedback. The atom number fluctuations in the "sine" and "cosine" BEC modes are even larger than that of the scattered photon number.    

The quantum simulation results show that the electronic feedback loop based on the photodetection can be applied to control the excitation of specific BEC modes without the introduction of additional noise in their quadratures. Although it has not been discussed in the paper, the use of electronic feedback can be more efficient and result in faster growth of BEC nonuniformity than the cavity-based feedback alone. This can happen since the electronic feedback strength can be made quite large in the beginning of the process. Thus an essential feedback action on the atoms can be obtained even for quite few scattered photons. These aspects, thought for slightly different system, have been addressed in ~\cite{ivanov-jetp-lett,ivanov-jetp}.         

\section{Acknowledgments}

The authors acknowledge Saint-Petersburg State University for a research grant 11.38.640.2013 and EPSRC Project EP/I004394/1.

\subsection{References}
\label{refs}

\end{document}